\documentstyle[amssym]{mn}

\font\japit = cmti10 at 10truept
\input{epsf.sty}

\title
     [Linear Redshift Distortions and Power in PSCz]
{\vglue-3.0truecm
\centerline{\japit Accepted for publication in Monthly Notices}
\vglue 2.5truecm
\noindent
Linear Redshift Distortions and Power in the PSCz Survey
\author[A. J. S. Hamilton, M. Tegmark and N. Padmanabhan]
     {A. J. S. Hamilton$^1$, Max Tegmark$^2$ and Nikhil Padmanabhan$^3$ \\
	$^1$JILA and Dept.\ Astrophysical \& Planetary Sciences,
	Box 440, U. Colorado, Boulder CO 80309, USA; \\
	\ Andrew.Hamilton@Colorado.EDU; http:$/\!/$casa.colorado.edu/$\sim$ajsh/ \\
	$^2$Dept. of Physics, Univ. of Pennsylvania, Philadelphia, PA 19104, USA;
	max@physics.upenn.edu; http:$/\!/$www.hep.upenn.edu/$\sim$max/ \\
	$^3$Department of Physics, Stanford University, CA 94305, USA; paddy@perseus.stanford.edu}
}

\newcommand{\rmn}{\rm}
\newcommand{\bmi}{\bmath}

\newcommand{\mx}{\bf}		

\newcommand{\be}{\begin{equation}}
\newcommand{\ee}{\end{equation}}
\newcommand{\ba}{\begin{eqnarray}}
\newcommand{\ea}{\end{eqnarray}}

\newcommand{\im}{{\rmn i}}	

\newcommand{\r}{{\bmi r}}
\newcommand{\bv}{{\bmi v}}

\newcommand{\bnabla}{{\bf\nabla}}

\newcommand{\Mpc}{{\rmn Mpc}}

\newcommand{\mxC}{{\mx C}}

\newcommand{\mxN}{{\mx N}}
\newcommand{\mxS}{{\mx S}}

\newcommand{\el}{\ell}

\newcommand{\aj}[2]{AJ, #1, #2}
\newcommand{\apj}[2]{ApJ, #1, #2}
\newcommand{\apjs}[2]{ApJS, #1, #2}

\newcommand{\mn}[2]{MNRAS, #1, #2}
\newcommand{\pr}[2]{Phys.\ Rev., #1, #2}

\newcommand{\xilcontsfig}{
    \begin{figure}
    \begin{center}
    \leavevmode
    \epsfxsize=3.3in	
    \epsfbox{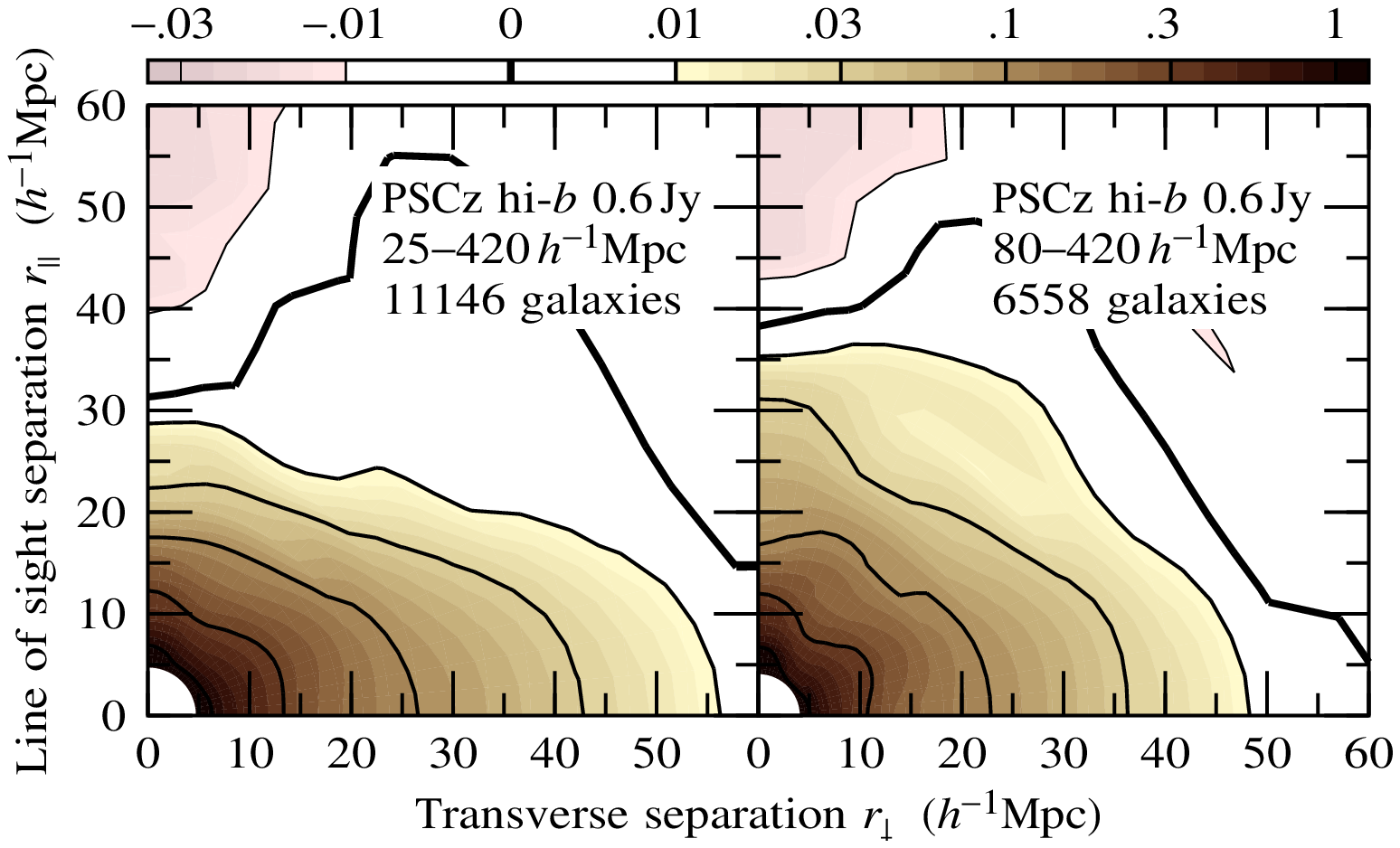}
    \end{center}
    \caption[1]{\small
Contour plots of the redshift space two-point correlation function
in the PSCz survey with the high galactic latitude angular mask,
(left) beyond $25 \, h^{-1} {\rm Mpc}$,
and (right) beyond $80 \, h^{-1} {\rm Mpc}$.
Thin, medium, and thick contours represent
negative, positive, and zero values respectively.
The correlation function has been smoothed over pair separation
$r = ( r_\perp^2 + r_\parallel^2 )^{1/2}$
with a tophat window of width $0.2$~dex,
and over angles $\theta = \tan^{-1} ( r_\perp / r_\parallel )$
to the line of sight with a Gaussian window
with a 1$\sigma$ width of $10^\circ$.
    \label{xilconts}
    }
    \end{figure}
}

\newcommand{\rhofig}{
    \begin{figure}
    \begin{center}
    \leavevmode
    \epsfxsize=2in	
    \epsfbox{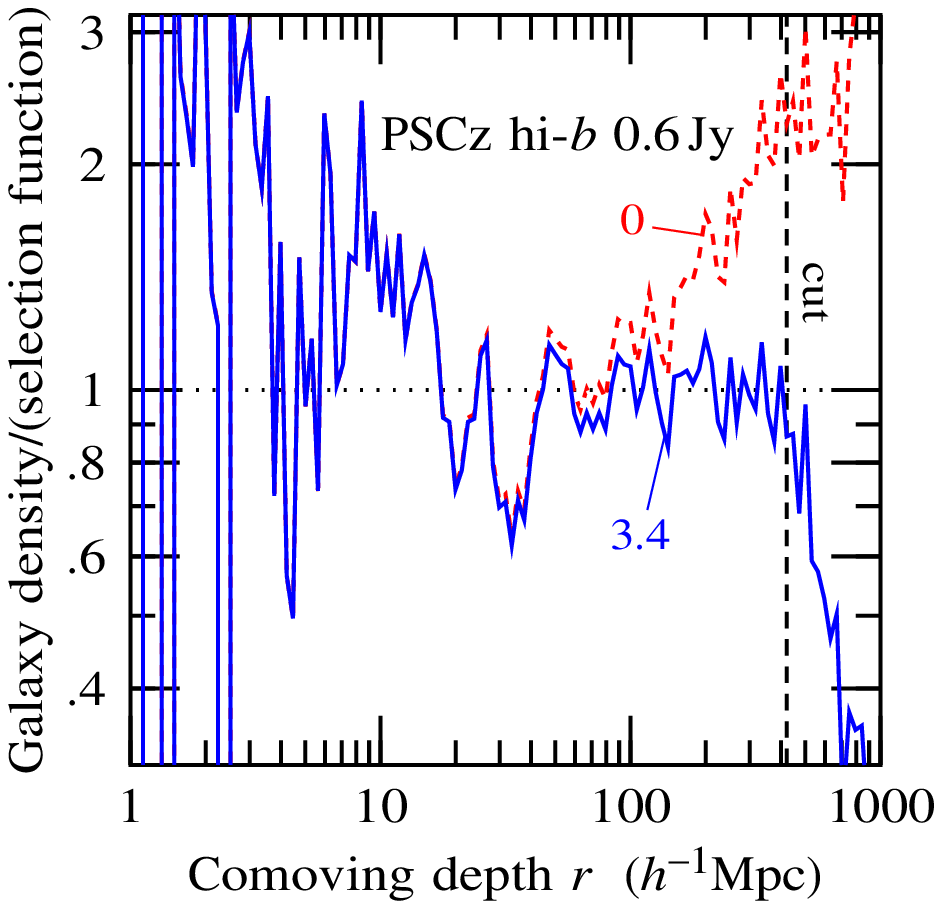}
    \end{center}
    \caption[1]{
Ratio of the observed galaxy number density to the maximum likelihood
selection function at radial depth $r$ in the PSCz survey,
averaged in depth bins $0.025$~dex wide.
The solid line assumes that galaxies evolve with
luminosity $\propto (1+z)^{3.4}$,
while the dashed line assumes no luminosity evolution.
For the analysis of this paper,
the survey is cut at a comoving depth of
$10^{2.625} \, h^{-1} \Mpc \approx 420 \, h^{-1} \Mpc$,
indicated by the vertical dashed line.
    \label{rho}
    }
    \end{figure}
}

\newcommand{\xikfig}{
    \begin{figure}
    \begin{center}
    \leavevmode
    \epsfxsize=2.3in	
    \epsfbox{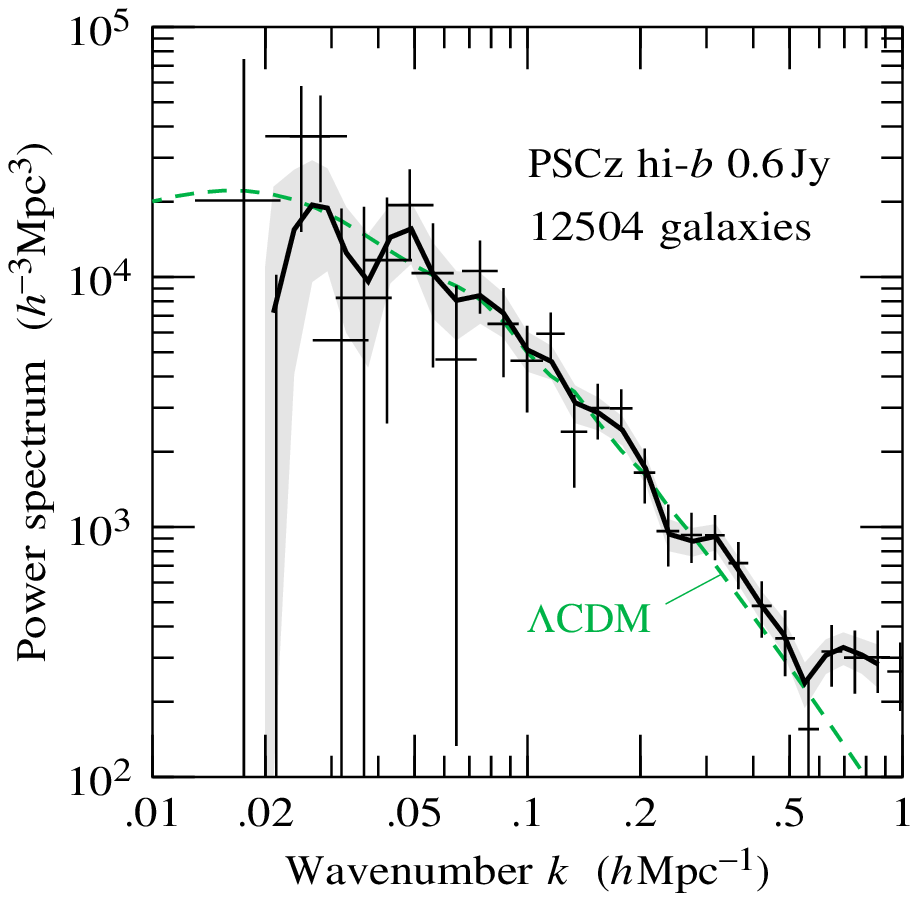}
    \end{center}
    \caption[1]{\small
Galaxy-galaxy power spectrum
measured from the PSCz 0.6~Jy survey with the high latitude angular mask.
This is the real space power spectrum,
the redshift distortions having been isolated into
galaxy-velocity and velocity-velocity power spectra.
The solid line is the correlated power spectrum,
and the shaded region its 1-$\sigma$ error,
while points with error bars constitute the decorrelated power spectrum
(Hamilton \& Tegmark 2000).
Each point of the decorrelated power spectrum is uncorrelated
with all other points.
The dashed line is a linear model power spectrum
from Eisenstein \& Hu (1998),
a COBE-normalized, untilted, flat $\Lambda$CDM model with
$\Omega_{\rm m} = 0.3$, $\Omega_\Lambda = 0.7$.
    \label{xik0}
    }
    \end{figure}
}

\newcommand{\betafig}{
    \begin{figure*}
    \begin{minipage}{175mm}
    \begin{center}
    \leavevmode
    \epsfxsize=5in	
    \epsfbox{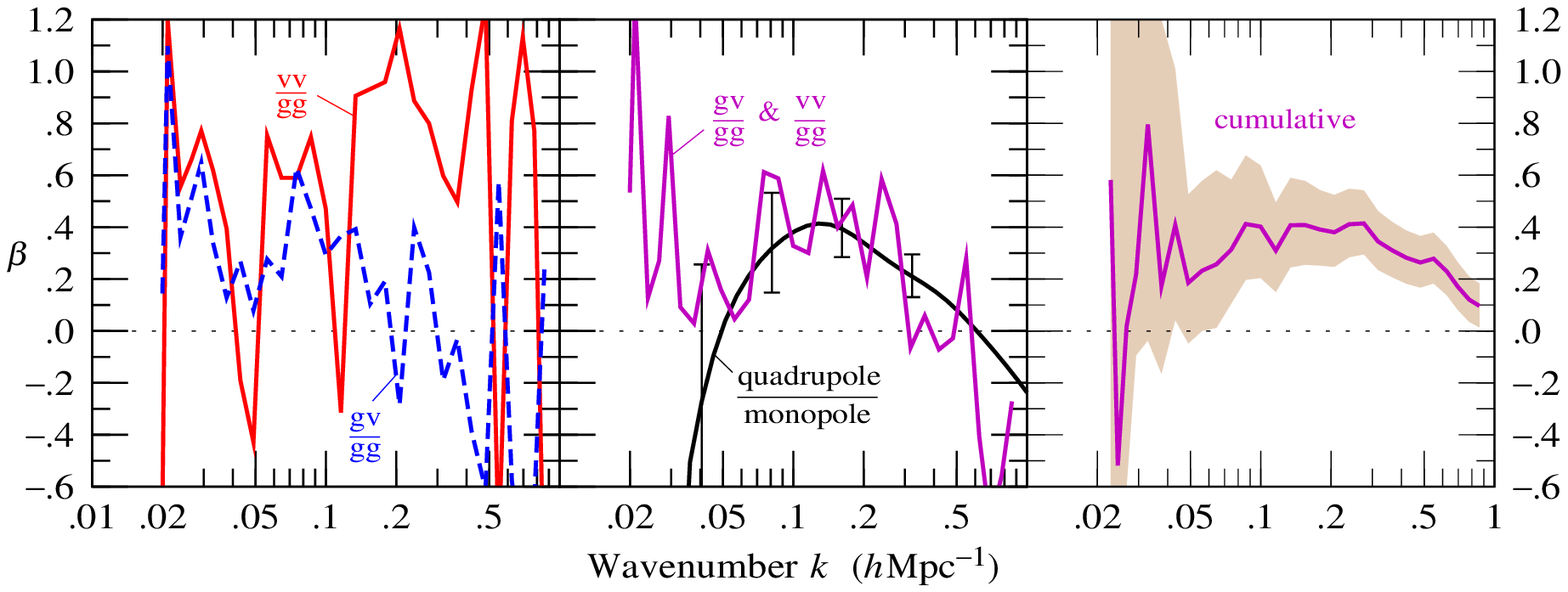}
    \end{center}
    \caption[1]{\small
Linear redshift distortion parameter $\beta \equiv f/b$
as a function of wavenumber $k$ in the PSCz 0.6~Jy survey.
(Left)
Dashed line is estimated
$r \beta$ from the ratio of galaxy-velocity to galaxy-galaxy power,
$P_{\rm gv}/P_{\rm gg}$.
Solid line is estimated
$\beta$ from the square root of the ratio of velocity-velocity to
galaxy-galaxy power,
$\mbox{sign}(P_{\rm vv}/P_{\rm gg}) | P_{\rm vv}/P_{\rm gg} |^{1/2}$.
(Middle)
Best fit value of $\beta$ from the combined ratios of
galaxy-velocity and velocity-velocity to galaxy-galaxy powers,
assuming unit galaxy-velocity correlation coefficient, $r = 1$.
For comparison, the graph also shows the value of $\beta$ inferred from
the ratio of quadrupole-to-monopole power,
computed in the manner of Hamilton (1995, 1998).
(Right)
Cumulative best fit value of $\beta$ including all data
at wavenumbers less than or equal to the wavenumber specified on the $x$-axis,
again assuming $r = 1$.
The shaded region is the 1-$\sigma$ uncertainty in $\beta$,
marginalized over the overall amplitude of the power spectrum.
    \label{beta}
    }
    \end{minipage}
    \end{figure*}
}

\begin{document}

\maketitle

\begin{abstract}
We present a state-of-the-art linear redshift distortion analysis of the
recently published {\it IRAS} Point Source Catalog Redshift Survey (PSCz).
The procedure involves linear compression into 4096 Karhunen-Lo\`eve
(signal-to-noise) modes
culled from a potential pool of $\sim 3 \times 10^5$ modes,
followed by quadratic compression into three separate power spectra,
the galaxy-galaxy, galaxy-velocity, and velocity-velocity power spectra.
Least squares fitting to the decorrelated power spectra
yields a linear redshift distortion parameter
$\beta \approx \Omega_{\rm m}^{0.6}/b = 0.41^{+0.13}_{-0.12}$.
\end{abstract}



\section{Introduction}
\label{intro}

The {\it IRAS\/} Point Source Catalog Redshift Survey (PSCz),
made public on 9 January 2000
(Saunders et al.\ 2000),
covers more volume of the Universe
than any other publically available redshift survey.
This,
together with the considerable care taken by its authors
to ensure uniformity of selection over a prescribed volume,
makes the PSCz the finest available galaxy survey
for measuring the power spectrum
and its redshift distortions at large, linear scales.

The purpose of this {\em Letter\/} is to report
a state-of-the-art linear redshift distortion analysis of the PSCz.
Our approach has its roots in the work of
Fisher, Scharf \& Lahav (1994),
who were the first to apply a maximum likelihood approach to the analysis
of linear redshift distortions,
and in the work of 
Heavens \& Taylor (1995, hereafter HT),
who may be credited with accomplishing the first linear likelihood
analysis designed to retain as much information as possible at the
largest, linear scales.
Fisher et al.\ and HT applied their approaches to the
the {\it IRAS} 1.2~Jy survey (Fisher et al.\ 1995),
obtaining values of the redshift distortion parameter
$\beta = 0.96^{+0.20}_{-0.18}$
and
$\beta = 1.0 \pm 0.3$
respectively,
where $\beta \equiv f/b$ is the dimensionless linear growth rate
$f \approx \Omega_{\rm m}^{0.6}$
divided by the linear galaxy-to-mass bias factor $b$.
A comprehensive review of these and other measurements of
linear redshift distortions, complete up to mid-1997,
may be found in Hamilton (1998).


Most recently, Tadros et al.\ (1999) applied HT's method to the PSCz survey,
conservatively cut to 0.75~Jy.
To deal with the more complicated PSCz mask,
Tadros et al.\ used an improved treatment of the angular mask,
allowing them to include monopole and dipole modes.
Tadros et al.\ obtained
$\beta = 0.58 \pm 0.26$
(marginalized error)
if the shape of the power spectrum was fixed to be that of a
$\Gamma = 0.2$ CDM model,
or
$\beta = 0.47 \pm 0.16$
(conditional error, with the power spectrum fixed at its best fit)
if the power spectrum was simultaneously measured in 6 bins.

The goal of the present paper is to measure the linear power spectrum
and its redshift distortions at high resolution with minimal loss of
information.
The paper brings together several ideas from the literature,
chief amongst which are:
(1) use of the logarithmic spherical wave basis
(Hamilton \& Culhane 1996);
(2) linear compression into Karhunen-Lo\`eve (KL), or signal-to-noise,
eigenmodes
(Bond 1995;   Bunn \& Sugiyama 1995; Vogeley \& Szalay 1996;
Matsubara, Szalay \& Landy 2000);
(3) quadratic compression into high resolution powers
(Tegmark 1997, 1998, hereafter T97, T98;
Tegmark, Taylor \& Heavens 1997, hereafter TTH;
Tegmark et al.\ 1998; Padmanabhan, Tegmark \& Hamilton 2000);
(4) inclusion of stochastic bias in the analysis
(Pen 1998;
Tegmark \& Peebles 1998;
Dekel \& Lahav 1999);
(5) decorrelation
(Hamilton 1997; Tegmark \& Hamilton 1998; Hamilton \& Tegmark 2000);
(6) the elimination of bias resulting from the `pair-integral constraint',
by the trick of making all KL modes orthogonal to the mean mode
(Fisher et al.\ 1993;
Tegmark et al.\ 1998).


\xilcontsfig


\section{The 2-point correlation function in redshift space}

Figure~\ref{xilconts}
shows the redshift space 2-point correlation function of PSCz.
The expected large scale squashing effect
caused by infall toward overdense regions is plainly visible.
Nonlinear fingers-of-god show up as
$\sim 4 \, h^{-1} {\rm Mpc}$ wide by $20 \, h^{-1} {\rm Mpc}$ long
enhancements along the line-of-sight axis.

There is little sign of the large scale `arm-of-god' affliction
beyond $80 \, h^{-1} \Mpc$ found by
Hamilton (1995)
in the QDOT survey,
the 1-in-6 precursor to PSCz
(Lawrence et al.\ 1999).
The contour plots in Figure~\ref{xilconts} are constructed in a manner
essentially identical to those in Hamilton (1995) and Hamilton (1998).
The only difference is that,
thanks to the lower noise level of the PSCz,
the contour plots in Figure~\ref{xilconts} extend to larger scales
and to lower contour levels than those of Hamilton (1995, 1998).

\section{Analysis}

The analysis will be described in full in a subsequent paper.
Here we summarize the main points.


As emphasized by Tadros et al.\ (1999),
in studies of weak clustering at large scales,
uniformity of selection is paramount,
since non-uniformity will masquerade as spurious excess power.
Following the advice of
Saunders et al.\ (2000)
and Tadros et al.,
we adopt the high-latitude angular mask
({\tt hibpsczmask.dat}, part of the PSCz package).



\rhofig

It is important to measure the radial selection function
of the survey as accurately as possible,
since errors in the selection function translate
into spurious large scale power.
We use the maximum likelihood method of
Sandage, Tammann \& Yahil (1979),
and fit the selection function
to a smooth analytic function
with enough free parameters to yield an excellent fit.
The assumed geometry is that of
a flat $\Omega_{\rm m} = 0.3$, $\Omega_\Lambda = 0.7$ Universe.
The measurement yields evidence for what appears to be strong evolution,
which we model as luminosity evolution with luminosity $\propto (1+z)^{3.4}$.
Similarly large evolution was previously reported in the QDOT survey
by Saunders et al.\ (1990).
Figure~\ref{rho}
shows the observed number density of galaxies,
divided by the measured selection function,
both with and without evolution.

Figure~\ref{rho}
also shows what appears to be incompleteness beyond $\sim 420 \, h^{-1} \Mpc$.
This may be presumed to be the incompleteness at high redshift
described in \S4.2 of Saunders et al.\ (2000),
associated with the policy
not to pursue redshifts of galaxies optically fainter than $b_J = 19.5^{\rm m}$.
Since this incompleteness is greater in regions of higher
optical extinction, and is systematic rather than random over the sky
(Fig~4. of Saunders et al.\ 2000),
we choose to cut the survey at
$10^{2.625} \, h^{-1} \Mpc \approx 420 \, h^{-1} \Mpc$.
%
We also choose to cut the survey at a near distance of $1 \, h^{-1} \Mpc$,
in order to eliminate the local nonlinear region,
the Local Group of galaxies.
This leaves 12504 galaxies in the sample.


As a working basis from which Karhunen-Lo\`eve modes
(see below)
are constructed,
we use logarithmic spherical waves (Hamilton \& Culhane 1996),
which are orthonormal eigenfunctions
$Z_{\omega \el m}(\r)
= (2 \pi)^{-1/2} \discretionary{}{}{} e^{- (3/2 + \im \omega) \ln r} \discretionary{}{}{} Y_{\el m} (\hat\r)$
of the complete set of commuting Hermitian operators
\begin{equation}
\label{op}
    \im \left( {\partial \over \partial \ln r} + {3 \over 2} \right) =
  - \im \left( {\partial \over \partial \ln k} + {3 \over 2} \right)
  \ , \quad
  L^2
  \ , \quad
  L_z
  \ .
\end{equation}
The advantages of this basis are:
(1) the linear redshift distortion operator is diagonal in this basis
(Hamilton \& Culhane 1996);
(2) radial modes can be transformed rapidly
between real, $\omega$, and Fourier space using Fast Fourier Transforms;
(3) the radial modes discretize naturally
on to a grid of radial depths $r$ that is uniformly spaced in the log,
which makes it well suited to surveys like PSCz,
where the flux limit causes the survey to be finely sampled nearby,
and coarsely sampled far away.

The principal numerical limitation to the linear likelihood method
is the number $N$ of modes that can be included in the Gaussian likelihood
function.
Solving the likelihood formulae involves manipulating
$N \times N$ matrices, which is a $\sim N^3$ process.
Thus the modes should be crafted so as to include as much information
as possible about the quantities of interest, the parameters to be measured
(HT; TTH).

Here we use Karhunen-Lo\`eve (KL), or signal-to-noise, modes,
as first suggested in the context of galaxy clustering by
Vogeley and Szalay (1996).
The covariance matrix
of galaxy densities
is a sum
$\mxC = \mxS + \mxN$
of cosmic signal $\mxS$ and Poisson sampling noise $\mxN$.
KL modes are constructed by diagonalizing the
signal-to-noise matrix $\mxN^{-1/2} \mxS \mxN^{-1/2}$,
and selecting the eigenmodes with the largest eigenvalues.


The main difficulty with the KL procedure is that
in order to construct $N$ modes it is necessary to diagonalize
a matrix of dimension $\gg N$.
But this would seem to defeat the original goal,
to avoid manipulating huge matrices.
We get around this difficulty by constructing KL modes in a two stage process,
constructing first angular KL modes,
then radial KL modes within each angular mode.
The resulting KL modes are not perfect,
since they ignore covariances between radial modes
belonging to different angular modes.
But the KL modes need not be perfect;
it is enough that they should contain virtually all the information of interest.


In practice,
we use 1600 angular modes, and 192 radial modes,
so there is a potential pool
of $1600 \times 192 \approx 3 \times 10^5$ modes.
This is $75 \approx 4^3$ times the number 4096 of KL modes
retained in the analysis.
In effect,
the KL modes are spatially over-resolved by a factor of about 4
in each dimension.


In order to isolate the effects of the selection function,
and of the motion of the Local Group (LG)
through the Cosmic Microwave Background (CMB),
we choose the first four angular `KL' modes to be
the cut (i.e. masked) monopole,
and three cut dipole modes.
The cut dipole modes contain admixtures of the cut monopole,
to make them orthogonal to the latter,
and all other angular KL modes are orthogonal to the cut monopole and dipole.
Within the cut monopole mode,
the first two radial modes are the mean (selection function) mode
and the Local Group mode
(equation~4.42 of Hamilton 1998),
the latter containing a small admixture of the mean to make it
orthogonal to the cut (i.e. radially masked) mean.
Within the cut dipole modes,
the first radial mode is the Local Group mode.

The mean mode is used in determining the maximum likelihood
normalization of the selection function,
but is then discarded from the analysis,
since it is impossible to measure the fluctuation of the mean mode.

Since all KL modes are orthogonal to the mean,
there is no need to subtract the mean.
Making the KL modes orthogonal to the mean immunizes their amplitudes
against uncertainty in the mean density,
and ensures that the measurement of power is unbiased by
the so-called `pair-integral constraint'.
This trick was first used by Fisher et al.\ (1993).


The fact that the selection function is measured in redshift space, not real
space, leads to a correction to the linear redshift distortion operator
for the monopole modes, described in \S4.4 of Hamilton (1998).
The correction leads to a noticeable reduction in measured power
at the largest scales.


The motion of the LG through the CMB is known
(Courteau \& van den Bergh 1999;
Lineweaver et al.\ 1996).
We correct
the amplitudes of the four LG modes (one cut monopole and three cut dipole)
for this motion,
and include them in the analysis,
as did Tadros et al.\ (1999).



The amplitudes of the 4096 KL modes constitute the `data'
which now feed into the Gaussian likelihood function.
Rather than attempting to solve the likelihood function directly
for values of cosmological parameters,
we instead quadratically compress
(T97; TTH)
the data into an intermediate number
(here 147, plus one for the normalization of the selection function)
of parameters that contain essentially all the information of interest.
The advantage of quadratic compression is that it
allows large numbers of parameters to be measured rapidly,
without a time-consuming nonlinear search for an extremum in a
high-dimensional space.
Key to the quadratic method
is that the intermediate parameters
should be linearly related to the mean and covariance.
%
%
In a linear redshift distortion analysis,
the intermediate parameters that emerge naturally
are the galaxy-galaxy, galaxy-velocity, and velocity-velocity power spectra
(Kolatt \& Dekel 1997;
Tegmark 1998;
Pen 1998;
Tegmark \& Peebles 1998;
Dekel \& Lahav 1999):
\begin{equation}
\label{Pks}
  \begin{array}{r@{\ }c@{\ }c@{\ }c@{\ }l}
    \mbox{galaxy-galaxy power} & : & P_{\rm gg}(k) & = & b(k)^2 P(k) \\
    \mbox{galaxy-velocity power} & : & P_{\rm gv}(k) & = & r(k) b(k) f P(k) \\
    \mbox{velocity-velocity power} & : & P_{\rm vv}(k) & = & f^2 P(k) \\
  \end{array}
\end{equation}
where $b(k)$ is the (possibly scale-dependent) galaxy-to-mass bias factor,
$r(k) \in [-1,1]$ is a (possibly scale-dependent)
galaxy-velocity correlation coefficient,
$f \approx \Omega_{\rm m}^{0.6}$ is the
dimensionless linear growth rate,
and $P(k)$ is the mass power spectrum.
More correctly, the `velocity' here refers to minus the velocity divergence,
which in linear theory is related to the mass (not galaxy)
overdensity $\delta$ by
$f \delta + \bnabla \cdot \bv = 0$,
where $\bnabla$ denotes the comoving gradient in velocity units.


Although the three power spectra are interesting in their own right,
here we report results in terms of conventional parameters:
the linear redshift distortion parameter $\beta \equiv f/b$,
the galaxy-velocity correlation coefficient $r$,
and the galaxy-galaxy power spectrum $P_{\rm gg}(k)$.

The estimate of power that emerges most directly from quadratic compression
is the power spectrum convolved with the Fisher matrix
(T97; TTH).
Physically, this represents the power spectrum
convolved with the Fourier transform of the optimally weighted survey window.
We call this estimate, appropriately renormalized,
the `correlated' power, since the errors are correlated.
Deconvolving the correlated power yields the `raw' power spectrum.
The raw power tends to be noisy and anti-correlated, and is not very useful.
Half way between the correlated and raw power spectra is the `decorrelated'
power spectrum, which is partially deconvolved in such a way that
estimates of power at different wavenumbers are uncorrelated with each other
(Hamilton 1997; Hamilton \& Tegmark 2000).

The quadratic method requires that prior powers
$P_{\rm gg}(k)$,
$P_{\rm gv}(k)$,
and
$P_{\rm vv}(k)$
be specified.
The maximum likelihood (ML) solution
is that power spectrum for which the estimated power equals the prior power.
The ML solution can be obtained by folding the estimate back
into the prior and iterating to convergence
(Bond, Jaffe \& Knox 1998).

However, in this paper the power is permitted to be an arbitrary
function of wavenumber, with no penalty against violently varying power spectra.
To express our Bayesian prejudice in favour of a smoothly varying power
spectrum, we choose to fold back into the prior not the raw
power spectrum,
but rather the correlated power spectrum, which is smooth(ish).
Moreover we iterate only once,
since there is no point in attempting
to overfit the bumps and wiggles in the power.
As the initial prior power,
we choose a COBE-normalized
flat $\Lambda$CDM model with $\Omega_{\rm m} = 0.3$, $\Omega_\Lambda = 0.7$
from Eisenstein \& Hu (1998).

Further, we set the redshift distortion parameter
in the prior power to a constant,
$\beta = 0.5$,
reflecting our Bayesian prejudice that this parameter
should be constant in the linear regime.
The data do not constrain the galaxy-velocity correlation coefficient
$r$ tightly, but are consistent with $r \approx 1$ at linear scales,
so we fix $r = 1$ in the prior.

\xikfig

\betafig

\section{Results}
\label{results}


Figure~\ref{xik0}
shows the galaxy-galaxy power spectrum measured from the PSCz 0.6~Jy survey.
This is the real space power spectrum,
the redshift distortions having been isolated into
the galaxy-velocity and velocity-velocity power spectra.
At nonlinear scales, $k \ga 0.3 \, h \, \Mpc^{-1}$,
the interpretation as a real space power spectrum becomes suspect,
since the extent to which nonlinear redshift distortions
can be approximated as ($k$-dependent) linear combinations
of the linear galaxy-velocity and velocity-velocity powers is unknown.

Figure~\ref{xik0}
also shows a linear model power spectrum from Eisenstein \& Hu (1998),
an observationally concordant COBE-normalized, untilted,
flat $\Lambda$CDM model with $\Omega_{\rm m} = 0.3$, $\Omega_\Lambda = 0.7$,
baryonic content $\Omega_b h^2 = 0.02$,
and Hubble constant
$h = 0.65$.
Interestingly,
the $\Lambda$CDM power spectrum fits the PSCz data well at linear scales,
$k \la 0.3 \, h \, \Mpc^{-1}$,
with no bias, $b = 1$.
The enhancement of the observed power over the model power at smaller scales,
$k \ga 0.5 \, h \, \Mpc^{-1}$,
can be attributed to nonlinearity
(Peacock \& Dodds 1996).

The real space power spectrum measured here appears consistent with the
redshift space power spectrum measured from PSCz
by Sutherland et al.\ (1999).


Figure~\ref{beta}
shows the redshift distortion parameter $\beta \equiv f/b$
measured as a function of wavenumber $k$.
The left panel shows the values of $\beta$
measured separately
from the ratio
$P_{\rm gv}(k)/P_{\rm gg}(k)$
of the galaxy-velocity to galaxy-galaxy power
on the assumption of unit galaxy-velocity correlation coefficient, $r = 1$,
and from the ratio
$\mbox{sign}(P_{\rm vv}/P_{\rm gg}) \left| P_{\rm vv}/P_{\rm gg} \right|^{1/2}$
of the square root of velocity-velocity to galaxy-galaxy power,
equations~(\ref{Pks}).
These are ratios of correlated power spectra,
since the decorrelated galaxy-velocity and velocity-velocity powers
are confusingly noisy.

The central panel of Figure~\ref{beta} shows the best fit value
of $\beta$ that comes from combining the two ratios of
galaxy-velocity and velocity-velocity to galaxy-galaxy power
at each wavenumber $k$.
Also shown for comparison is the value of $\beta$ inferred from
the ratio of smoothed quadrupole power to smoothed monopole power,
measured in the fashion of Hamilton (1995, 1998).
The two measures of $\beta$ agree at scales
$k \ga 0.04 \, h \, \Mpc^{-1}$.
The difference at
$k \la 0.04 \, h \, \Mpc^{-1}$
may plausibly be attributed
to the inability of the quadupole-to-monopole ratio to probe the largest scales.

The right panel of Figure~\ref{beta}
shows the best fit value of $\beta$ that emerges
from a least squares fit to the decorrelated powers
including data cumulatively up to the wavenumber specified on the $x$-axis,
again on the assumption of unit galaxy-velocity correlation coefficient,
$r = 1$.
The shaded region bounds the 1-$\sigma$ uncertainty in $\beta$,
where $\chi^2$ exceeds its minimum value by 1.

In carrying out the fit we fix the shape of the galaxy-galaxy power spectrum
to be that of the measured correlated power spectrum shown
in Figure~\ref{xik0}, but we allow the overall amplitude of the power
spectrum to vary.
The 1-$\sigma$ uncertainty in Figure~\ref{beta} is the uncertainty
marginalized over the amplitude of the power spectrum.
We also tried allowing the shape of the power spectrum to vary,
but the shape is already measured about as well as can be,
and we could not thereby reduce $\chi^2$ in a statistically significant way
(reduction by at least 1 for each additional degree of freedom).

The fit is consistent with constant $\beta$ for $k \la 0.3 \, h \, \Mpc^{-1}$,
and as an overall best fit we quote the value
$\beta = 0.41^{+0.13}_{-0.12}$
at $k = 0.274 \, h \, \Mpc^{-1}$.
$\Lambda$CDM with $\Omega_{\rm m} = 0.3$ and $\Omega_\Lambda = 0.7$
predicts $f = 0.513$,
consistent with our best fit result if {\it IRAS\/} galaxies are unbiased
or mildly biased.


We also tried allowing the galaxy-velocity correlation coefficient $r$ to vary.
At linear scales, $k \la 0.15 \, h \, \Mpc^{-1}$,
the best fit value is of order unity, $r \approx 1$,
as suggested by the approximate agreement in the two curves
in the left panel of Figure~\ref{beta},
but the uncertainty is large.
At smaller scales, $k \ga 0.15 \, h \, \Mpc^{-1}$,
the measured coefficient decreases below 1.
This is almost certainly caused by nonlinearity,
not by stochasticity in the linear galaxy-velocity correlation.
In fact fingers-of-god,
which appear extended along the line-of-sight in redshift space
rather than compressed,
are expected to cause the galaxy-velocity power to go negative at nonlinear
scales, yielding a negative galaxy-velocity correlation coefficient $r$.
The left panel of Figure~\ref{beta}
shows that the galaxy-velocity power indeed goes negative at
$k \ga 0.2 \, h \, \Mpc^{-1}$.

\section{Conclusions}

This {\it Letter\/} reports
a measurement of the linear power spectrum and its redshift distortions
in the {\it IRAS} PSCz 0.6~Jy survey
(Saunders et al.\ 2000)
using state-of-the-art methods,
including logarithmic wave expansion,
Karhunen-Lo\`eve compression,
quadratic compression,
decorrelation,
and isolation of the mean mode
in order to eliminate large scale bias from the `pair-integral-constraint'.

The measurement yields three separate linear power spectra,
the galaxy-galaxy, galaxy-velocity, and velocity-velocity power spectra.

The best fit overall value of the linear redshift distortion parameter,
$\beta \equiv f/b = 0.41^{+0.13}_{-0.12}$,
is consistent with, albeit slightly lower than,
that reported by Tadros et al.\ (1999)
for a more conservative 0.75~Jy cut of the survey.

The galaxy-velocity correlation coefficient $r$ is not tightly
constrained by the data, but is consistent with being about unity
at linear scales, $r \approx 1$.

The inferred real space power spectrum and its redshift distortions
are consistent with a COBE-normalized, untilted, flat $\Lambda$CDM
model with $\Omega_{\rm m} = 0.3$ and $\Omega_\Lambda = 0.7$,
if {\it IRAS\/} galaxies are an unbiased tracer of mass
at large scales in the Universe.

\section*{Acknowledgements}

This work was supported by
NASA ATP grant NAG5-7128
and by
NASA LTSA grant NAG5-6034.
AJSH thanks Michael Culhane for many conversations.

\end{document}